\documentstyle{rspublic}
\input{epsf.sty}

\newcommand{\Mpc}{\mbox{\,h$^{-1}$\,Mpc}}
\newcommand{\Mpccube}{\mbox{\,h$^{-3}$\,Mpc$^3$}}
\newcommand{\gs}
           {\mathrel{\hbox{\rlap{\hbox{\lower4pt\hbox{$\sim$}}}\hbox{$>$}}}}
\newcommand{\ls}
           {\mathrel{\hbox{\rlap{\hbox{\lower4pt\hbox{$\sim$}}}\hbox{$<$}}}}

\begin{document}

\title{First results from the 2dF galaxy \\ redshift survey}
\shorttitle{First results from the 2dF galaxy redshift survey}
\author{M\,a\,t\,t\,h\,e\,w C\,o\,l\,l\,e\,s\,s 
\thanks{{\sl On behalf of the 2dF Galaxy Redshift Survey Team:} 
Matthew Colless (MSSSO), Richard Ellis (IoA), Joss Bland-Hawthorn (AAO),
Russell Cannon (AAO), Shaun Cole (Durham), Chris Collins (LJMU), Warrick
Couch (UNSW), Gavin Dalton (Oxford), Simon Driver (UNSW), George
Efstathiou (IoA), Simon Folkes (IoA), Carlos Frenk (Durham), Karl
Glazebrook (AAO), Nick Kaiser (IfA), Ofer Lahav (IoA), Ian Lewis (AAO),
Stuart Lumsden (AAO), Steve Maddox (IoA), John Peacock (ROE), Bruce
Peterson (MSSSO), Ian Price (MSSSO), Will Sutherland (Oxford), Keith
Taylor (AAO)}}
\shortauthor{Matthew Colless}
\affiliation{Mount Stromlo and Siding Spring Observatories,\\
             The Australian National University}
\maketitle

\abstract The 2dF Galaxy Redshift Survey is a major new initiative to
map a representative volume of the universe. The survey makes use of the
2dF multi-fibre spectrograph at the Anglo-Australian Telescope to
measure redshifts for over 250,000 galaxies brighter than b$_J$=19.5 and
a further 10,000 galaxies brighter than R=21. The main goals of the
survey are to characterise the large-scale structure of the universe and
quantify the properties of the galaxy population at low redshifts. This
paper describes the design of the survey and presents some preliminary
results from the first 8000 galaxy redshifts to be measured.
\endabstract

\section{The goals of the survey}

Cosmology is undergoing a {\it fin-de-millennium} flowering brought on by
the hothouse of recent technological progress. Cosmography, one of the
roots of cosmology, is thriving too---maps of the distribution of
luminous objects are covering larger volumes, reaching higher redshifts
and encompassing a wider variety of sources than ever before. New ways
of interpreting these observations are yielding a richer and more
detailed picture of the structure and evolution of the universe, and of
the underlying cosmology. 

The 2dF Galaxy Redshift Survey at the Anglo-Australian Telescope aims to
map the optically-luminous galaxies over a statistically-representative
volume of the universe in order to characterise as fully as possible the
large-scale structure of the galaxy distribution and the interplay
between this structure and the properties of the galaxies themselves. In
doing so, the survey will address a variety of fundamental problems in
galaxy formation and cosmology. The major scientific goals include:

\begin{enumerate}
\item 
Measuring the power spectrum of the galaxy distribution on scales up to
a few hundred Mpc, filling the gap in our knowledge of the power
spectrum between scales less than 100\Mpc\ derived from earlier galaxy
redshift surveys and scales greater than 1000\Mpc\ probed by microwave
background anisotropy observations. The shape of the power spectrum on
these large scales provides a strong constraint on the nature of the
dark matter (i.e.\ whether it is hot or cold) and the total mass density
$\Omega$.
\item
Measuring the distortion of the large-scale spatial clustering pattern
in redshift space due to the peculiar velocity field produced by the
mass distribution. This distortion depends on both the mass density
parameter $\Omega$ and the bias factor $b$ of the galaxy distribution
with respect to the mass distribution, and its measurement constrains
the combination $\beta \approx \Omega^{0.6}/b$.
\item
Investigating the morphology of galaxy clustering and the statistical
properties of the fluctuations: on large scales, to determine whether
the density distribution is Gaussian as predicted by most inflationary
models of the early universe; on small scales, to probe the non-linear
evolution of the density field.
\item
Determining the variations in the spatial and velocity distributions of
galaxies as a function of galaxy properties, including luminosity and
spectral type. This provides a detailed picture of the link between the
masses and star-formation histories of galaxies and their local
environment within the large-scale distribution. Such information will
constrain models of galaxy formation and evolution.
\item
Tracking the galaxy luminosity function, clustering amplitude and mean
star formation rate out to a redshift of $z$$\sim$0.5, in order to test
models for the evolution of the galaxies' stellar populations and
large-scale distribution.
\item
Defining large, homogeneous samples of groups and clusters of galaxies
and other large-scale structures. Redshifts for many galaxies in a
representative sample of clusters and groups will give a snapshot of the
dynamical evolution of the most massive bound objects in the universe at
the present epoch. Mapping the infall patterns around clusters will also
yield dynamical estimates of cluster masses at large radii.
\end{enumerate}

The redshift survey will also provide a massive database for use in
conjunction with other surveys or as a source for follow-up programs.
Various interesting subsamples of galaxies (such as AGN or cDs) can be
defined using the positions, luminosities and spectral types provided by
the survey. Other rich veins of information can be tapped by correlating
the galaxies in this survey with sources found at other wavelengths
(X-ray, infrared, radio) by existing or planned satellite or
ground-based surveys (ROSAT, IRAS, WIRE, FIRST, etc.).

\section{Survey design}

The survey design seeks to achieve the above goals using the
capabilities of the 2dF multi-fibre spectrograph on the Anglo-Australian
Telescope (AAT) in approximately 100 nights of telescope time. A full
description of the 2dF facility is available on the WWW at {\tt
http://www.aao.gov.au/2df/}. However for the purposes of the survey, the
essential features of 2dF are its 2~degree diameter field of view
covered by 400 fibres and that it is attached to a 4m telescope. With
both the goals of the survey and the capabilities of 2dF in mind, and
with the limitation that the survey should not require more than about
100 nights of AAT time, we have arrived at the following survey design.

\subsection{Source catalogue}

\begin{figure}
\centering
\vspace*{5pt}
\parbox{\textwidth}{\epsfxsize=\textwidth \epsfbox{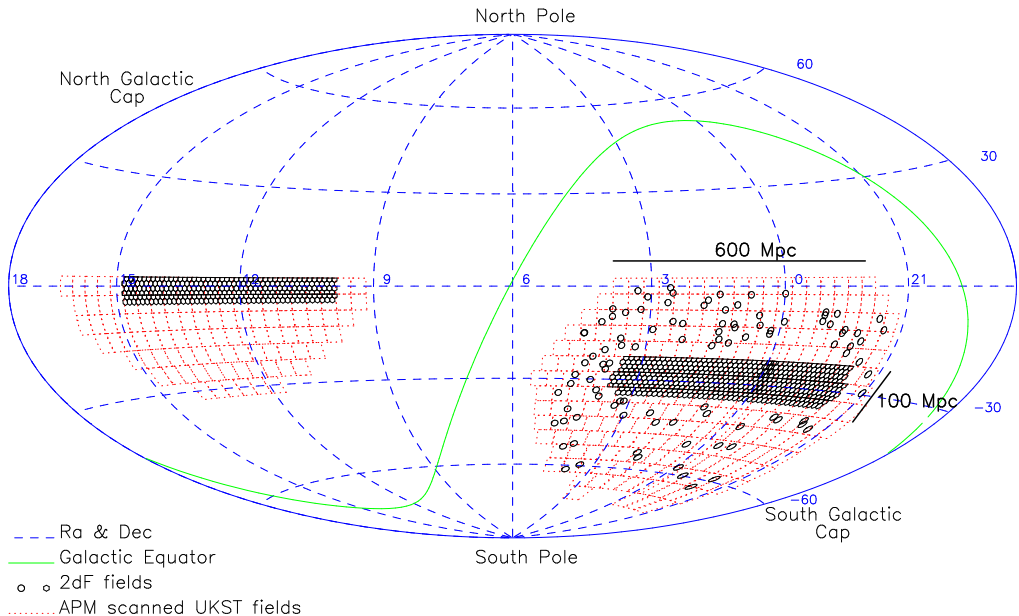}}
\vspace*{5pt}
\caption{The 2dF survey fields (small circles) superimposed on the APM
survey area (dotted outlines of Sky Survey plates). There are
approximately 140,000 galaxies in the 75$^\circ$$\times$15$^\circ$
South Galactic Hemisphere strip centred on the South Galactic
Pole, 70,000 galaxies in the 75$^\circ$$\times$7.5$^\circ$ North
Galactic Hemisphere equatorial strip, and 40,000 galaxies in the
100 random 2dF fields scattered over the entire southern region of the
APM galaxy survey. 
\label{fig:skyplot}}
\end{figure}

The source catalogue for the survey is a revised and extended version of
the APM galaxy catalogue (Maddox et~al.\ 1990a,b,c). This catalogue is
based on Automated Plate Measuring machine (APM) scans of 390 plates
from the UK Schmidt Telescope (UKST) Southern Sky Survey. The magnitude
system for the Southern Sky Survey is defined by the response of Kodak
IIIaJ emulsion in combination with a GG395 filter, zeropointed by CCD
photometry to the Johnson B band. The extended version of the APM
catalogue includes over 5 million galaxies down to b$_J$=20.5 in both
north and south Galactic hemispheres over a region of almost
10$^4$~sq.deg (bounded approximately by declination $\delta$$\leq$+3 and
Galactic latitude $b$$\gs$20). The fields included in the catalogue are
shown as the red squares in Figure~\ref{fig:skyplot}. The astrometry for
the galaxies in the catalogue has been significantly improved, so that
the rms error is now 0.25~arcsec for galaxies with b$_J$=17--19.5. Such
precision is required in order to minimize light losses with the
2~arcsec diameter fibres of 2dF. The photometry of the catalogue is
calibrated with numerous CCD sequences and has a precision of
approximately 0.2~mag for galaxies with b$_J$=17--19.5.

\subsection{Survey geometry}

The geometry of the survey was chosen to be an effective compromise
between the desire to sparsely sample the largest possible volume in
order to determine the power spectrum on very large scales and the
desire to fully sample a representative but compact volume in order to
investigate the redshift space distortions and the topology of the
galaxy distribution. There is also the observational requirement to
spread the survey over a wide R.A.\ range to permit efficient use of
telescope time. The survey geometry adopted is shown in
Figure~\ref{fig:skyplot}. It consists of two contiguous declination
strips plus 100 random 2-degree fields. One strip is in the southern
Galactic hemisphere and covers approximately
75$^\circ$$\times$15$^\circ$ centred close to the South Galactic Pole at
($\alpha$,$\delta$)=($01^h$,$-30$); the other strip is in the northern
Galactic hemisphere and covers 75$^\circ$$\times$7.5$^\circ$ centred at
($\alpha$,$\delta$)=($12.5^h$,$+00$). The 100 random fields are spread
uniformly over the 7000~sq.deg region of the APM catalogue in the
southern Galactic hemisphere. At the mean redshift of the surey
($\bar{z}\approx0.11$), 100\Mpc\ subtends about 20~degrees, so the two
strips are 375\Mpc\ long and have widths of 7.5\Mpc\ (south) and
3.8\Mpc\ (north). The volume directly sampled by the survey (out to
$z$=0.2) is 3$\times$10$^7$\Mpccube; the volume sparsely sampled
including the random fields is 1$\times$10$^8$\Mpccube.

\subsection{Sample selection}

The sample is chosen to be magnitude-limited at b$_J$=19.5 after
extinction-correcting all the magnitudes in the APM catalogue using the
extinction maps of Schlegel et~al.\ (1998). This limit was chosen
because the mean number of galaxies per square degree at b$_J$=19.5 is
well-matched to the density of fibres available with 2dF. Due to
clustering, however, the number in a given field varies considerably. To
make efficient use of the instrument we employ a sophisticated tiling
algorithm to cover the survey area with the minimum number of 2dF
fields. With this algorithm we are able to achieve a 93\% sampling rate
with on average fewer than 5\% wasted fibres per field. Over the whole
area of the survey there are in excess of 250,000 galaxies. The mean
redshift of this b$_J$=19.5 magnitude-selected sample is $\bar{z}$=0.11.

\subsection{The faint survey}

The `bright' survey described above will be supplemented by a `faint'
survey of 10,000 galaxies down to R=21. These are drawn from APM scans
of deep UK Schmidt Telescope films taken on Kodak TechPan emulsion. The
faint survey is limited to selected fields in the two survey strips and
is carried out as an over-ride on the bright survey in the best
observing conditions. The mean redshift of the faint survey is
$\bar{z}$=0.3, and thus extends the bright survey sample by a factor of
3 in depth and a factor of 10 in luminosity at the cost of a 10\%
increase in total observing time.

\section{Survey observations and status}

\begin{figure}
\centering 
\vspace*{5pt}
\parbox{\textwidth}{\epsfxsize=0.7\textwidth \epsfbox{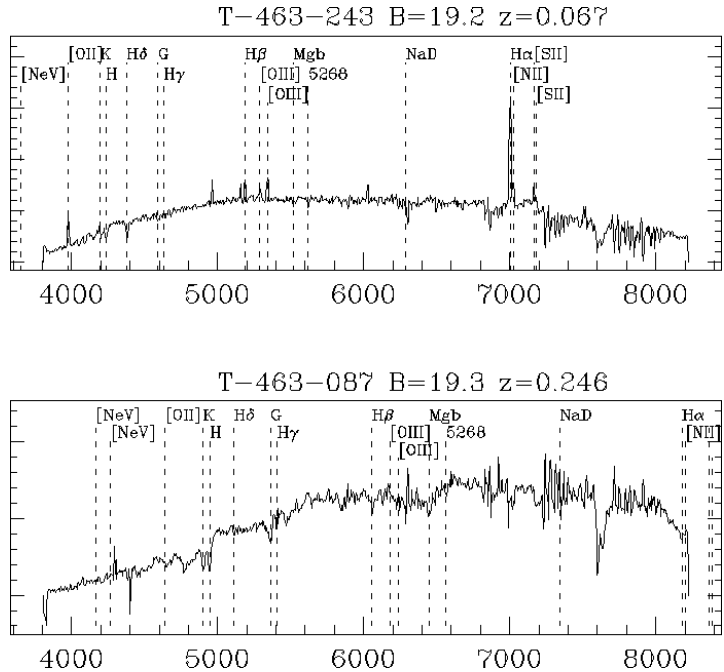}}
\vspace*{5pt}
\caption{Example spectra from the survey: a b$_J$=19.2 emission-line
galaxy at $z$=0.067 and a b$_J$=19.3 absorption-line galaxy at
$z$=0.246.
\label{fig:specs}}
\end{figure}

The total integration time on a given field has a lower limit set by the
time required for 2dF's robotic fibre positioner to configure the
off-line field-plate. Other than this hardware limitation, the
integration time is determined by the requirement that we can obtain
precise and reliable redshifts and spectral type classifications, which
is comfortably met with integrations of 60~min. The spectra we obtain
cover the range 3800--8000\AA\ with a two-pixel resolution of 8.6\AA,
and have a minimum S/N of about 10 per pixel at 5000\AA; most spectra
will easily exceed this target. Two example spectra from the survey are
shown in Figure~\ref{fig:specs}. One is a b$_J$=19.2 emission-line
galaxy at $z$=0.067 and the other is a b$_J$=19.3 absorption-line galaxy
at $z$=0.246.

This minimum S/N permits very reliable automatic spectral classification
and redshift measurement. Employing both a standard cross-correlation
and line-fitting code and a new code which uses principal component
analysis and $\chi^2$-fitting to simultaneously classify the spectrum
and measure its redshift (Glazebrook et~al.\ 1998), we find we achieve a
very high level of reliability. A comparison of the redshifts obtained
from these codes with redshifts determined via visual inspection shows a
very low level of failures in the automatic algorithms. The success rate
in identifying redshifts for survey galaxies is currently 90--95\%; the
goal is to achieve an overall success rate in excess of 95\%.

The first data for the survey was taken at the start of 2dF's
commissioning in November 1996. The first survey observations with all
400 fibres were obtained in October 1997. As of this writing (March
1998) we have obtained $\sim$8000 redshifts for the survey in 43
different fields (many of which were observed with only 200 fibres and
most of which were shared with the QSO survey---see Boyle, these
proceedings). At present the fibre positioner is taking 110~min to
configure a typical field, limiting the survey to 4--5 fields per night.
However it is confidently expected that fine-tuning of the robot's
hardware and software will reduce this configuration time to about
60~min within the next few months, doubling the rate at which fields are
done. We therefore currently expect to complete the survey before the
end of 2000.

\begin{figure}
\centering 
\vspace*{5pt}
\parbox{\textwidth}{(a)\epsfxsize=0.95\textwidth \epsfbox{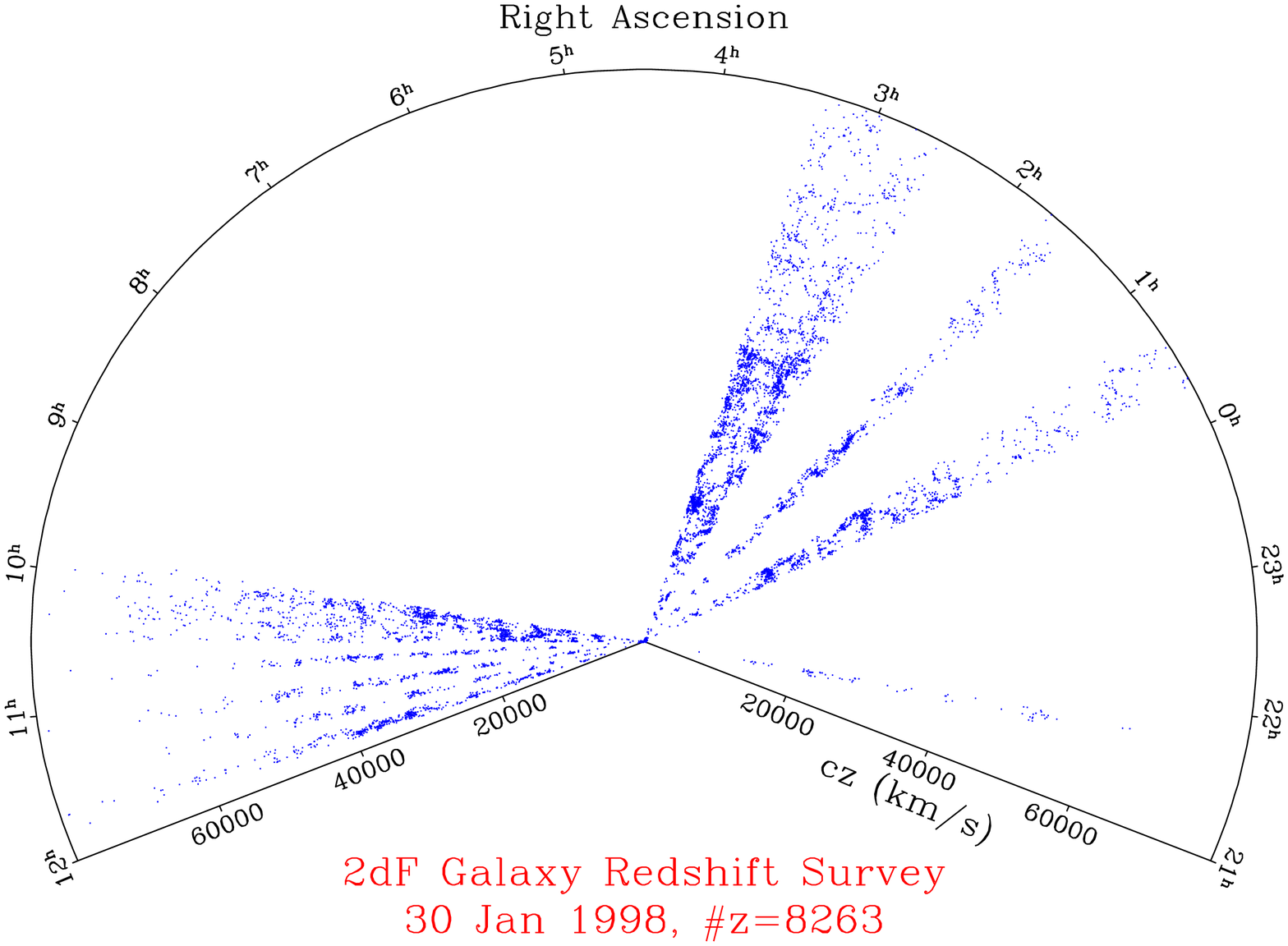}}\\
\vspace*{5pt}
\parbox{\textwidth}{(b)\epsfxsize=0.95\textwidth \epsfbox{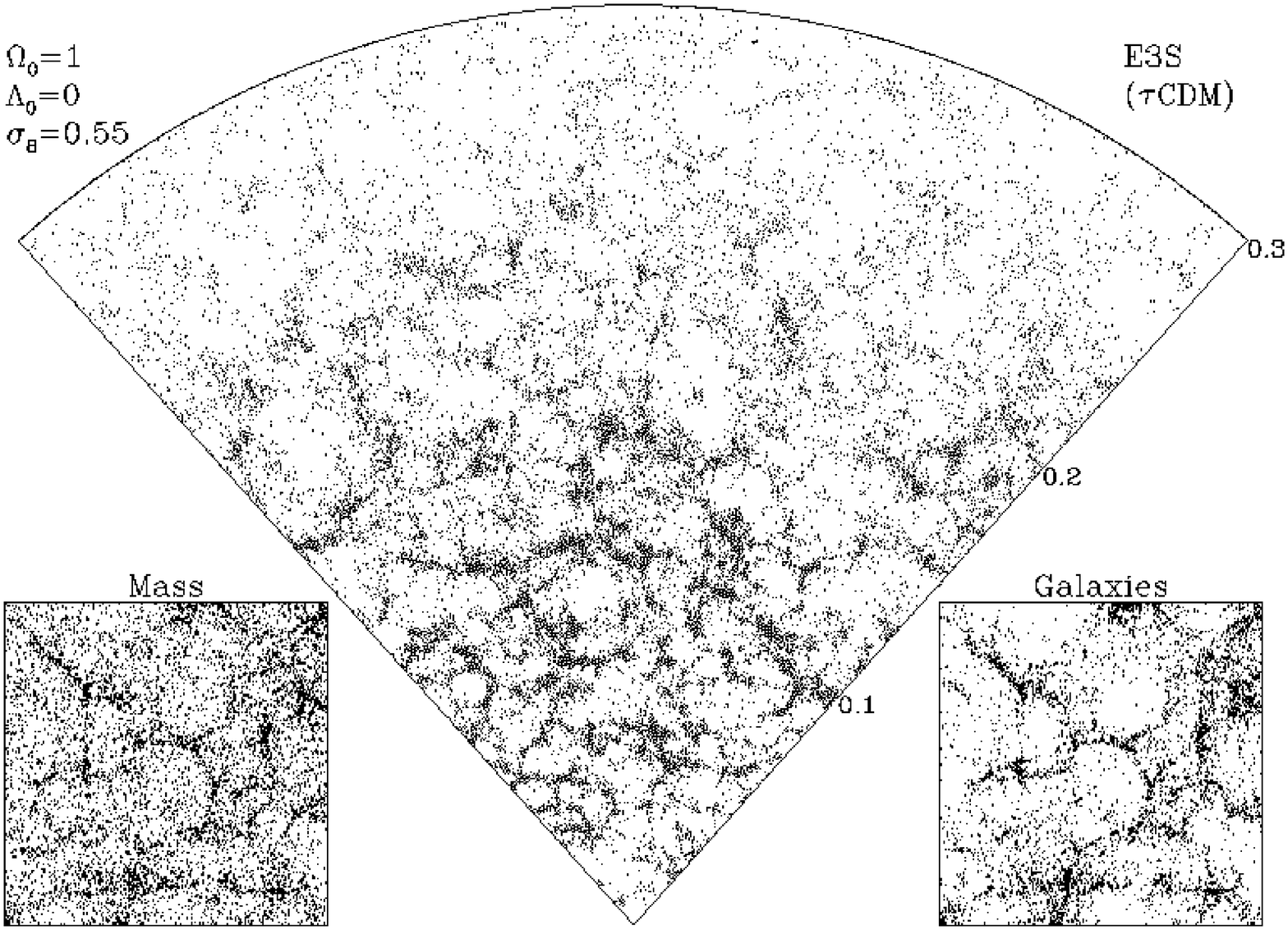}}
\vspace*{5pt}
\caption{(a)~A redshift slice for the galaxies observed to date,
combining northern and southern strips and including $\sim$8000 galaxies
($\sim$3\% of the full sample). (b)~A cone plot for a mock 2dF redshift
survey from Cole et~al.\ (1998).
\label{fig:zslice}}
\end{figure}

\section{Preliminary Results}

The survey is still in its infancy, with less than 3\% of the total
number of redshifts measured to date. Moreover the fields observed so
far are scattered over the survey volume, making the analysis of the
large-scale structure of the distribution problematic. Nonetheless, the
preliminary results presented here do provide some hints of the power
and scope of the full survey.

Figure~\ref{fig:zslice}a is a cone plot showing the distribution of the
8000 galaxies observed so far. Note that fields at all declinations are
projected onto this right ascension slice through the galaxy
distribution, which combines both the northern and southern strips. Even
with the highly incomplete filling of the slice by the fields obtained
to date, there are clear glimpses of significant large-scale structures.
To provide a visual reference for comparison, Figure~\ref{fig:zslice}b
shows a 90$^\circ$$\times$3$^\circ$ slice through a mock 2dF survey
based on a CDM N-body simulation and a recipe for galaxy biasing (see
Cole et~al.\ 1998).

\begin{figure}
\centering 
\vspace*{5pt}
\parbox{\textwidth}{\epsfxsize=0.8\textwidth \epsfbox{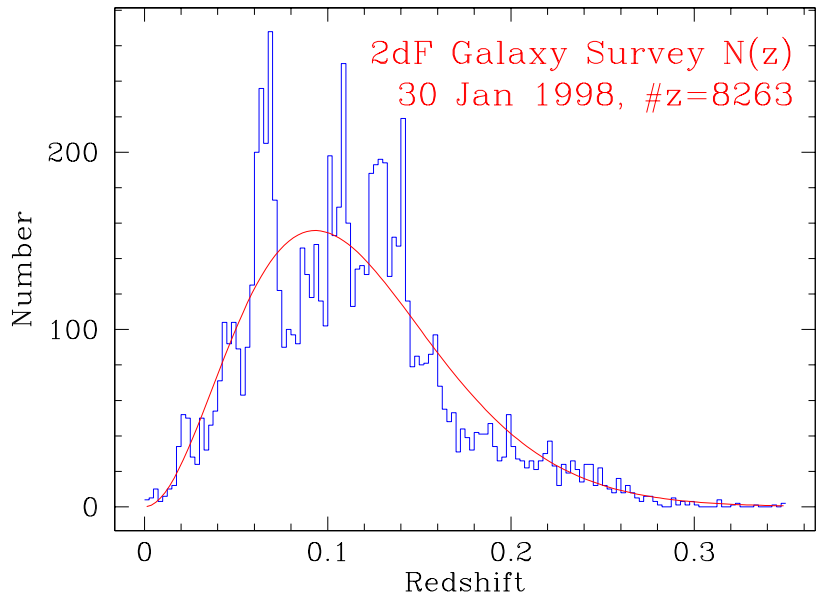}}
\vspace*{5pt}
\caption{A preliminary redshift distribution from the 2dF survey. The
smooth curve is the predicted redshift distribution neglecting
clustering.
\label{fig:nz}}
\end{figure}

Figure~\ref{fig:nz} shows the combined redshift distribution for all
fields in comparison to the predicted distribution in a homogeneous
universe. The mean redshift is 0.11, as expected, although the clear
signature of clustering shows that we are still far from having a
representative volume of the universe, in which the deviations from the
redshift distribution approach the Poisson limit.

\begin{figure}
\centering 
\vspace*{5pt}
\parbox{\textwidth}
  {\hspace*{0.1\textwidth}(a)~\epsfxsize=0.67\textwidth\epsfbox{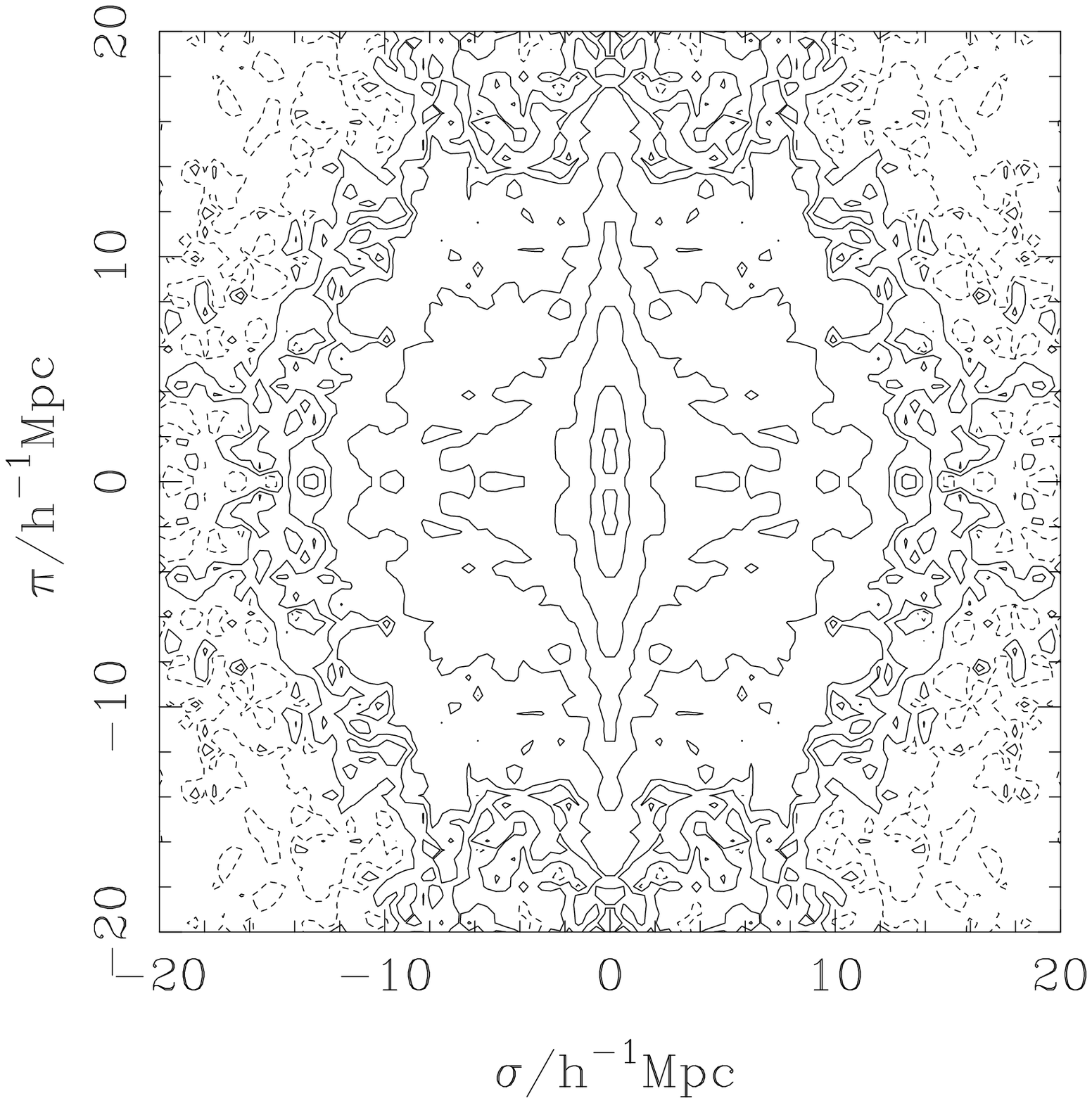}}\\
\vspace*{5pt}
\parbox{\textwidth}
  {\hspace*{0.1\textwidth}(b)~\epsfxsize=0.67\textwidth\epsfbox{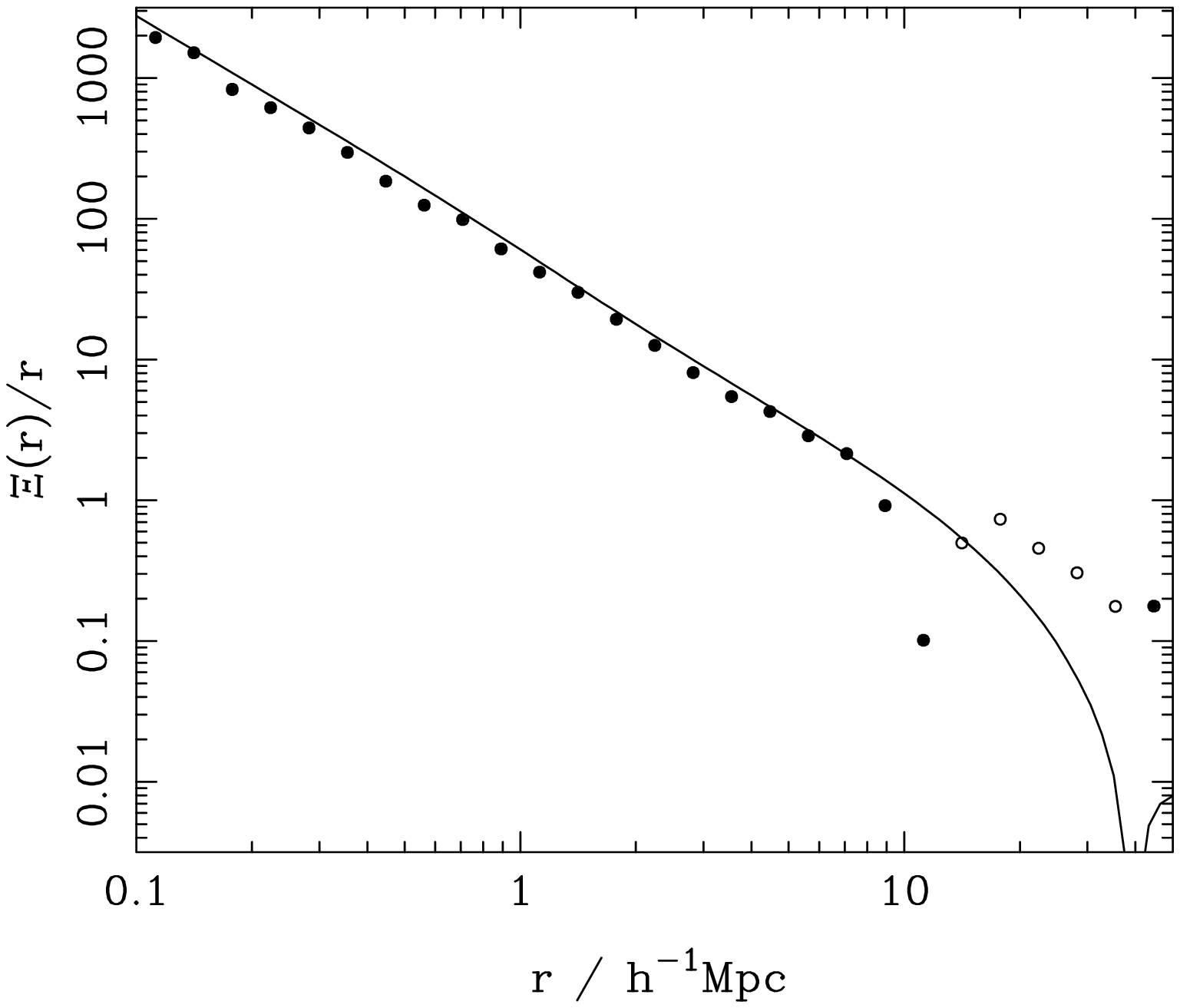}}
\vspace*{5pt}
\caption{(a)~The correlation function in redshift space
($\xi_s(\sigma,\pi)$) as a function of separation in the plane of the
sky ($\sigma$) and along the line-of-sight ($\pi$). The spatial
resolution of this contour plot is 0.5\,$h^{-1}$\,Mpc, and the contours
are $-$0.1, 0, 0.1, 0.2, 0.5, 1, 2, 5, 10, 30 and 50. (b)~A preliminary
estimate of the real-space galaxy correlation function. $\Xi(r)$ is the
projection of $\xi_s(\sigma,\pi)$ in the plane of the sky (i.e.\ onto
the $\sigma$ axis); $\Xi(r)/r \propto \xi_r(r)$ for a power-law
real-space correlation function $\xi_r(r)=(r/r_0)^{-\gamma}$. Solid and
open points indicate positive and negative values respectively. The
solid line shows the prediction from the deprojected power spectrum of
Baugh and Efstathiou (1993).
\label{fig:xiplots}}
\end{figure}

Preliminary measurements of the galaxy correlation function are shown in
Figure~\ref{fig:xiplots}. The redshift-space distortion of the
distribution is illustrated in Figure~\ref{fig:xiplots}a, which shows
$\xi_s(\sigma,\pi)$, the correlation function in redshift-space as a
function of separation in the plane of the sky ($\sigma$) and along the
line-of-sight ($\pi$). At small separations we see that the correlation
function is flattened along the line of sight due to the finger-of-god
effect. At sufficiently large separations (i.e.\ in the linear regime)
we expect a flattening in the plane of the sky that depends on
$\beta=\Omega^{0.6}/b$, although the preliminary determination of
$\xi_s(\sigma,\pi)$ presented here is too noisy on large scales to
permit an estimate of $\beta$. We can project $\xi_s(\sigma,\pi)$ in the
plane of the sky (i.e.\ onto the $\sigma$ axis) to obtain the projected
real-space correlation function $\Xi(r)$. Figure~\ref{fig:xiplots}b
shows $\Xi(r)/r$, since $\Xi(r)/r \propto \xi_r(r)$ (the deprojected
real-space correlation function) in the case of a power-law,
$\xi_r(r)=(r/r_0)^{-\gamma}$. For $r$$<$10\Mpc\ we find that $\Xi(r)/r
\propto r^{-1.7}$ in agreement with most previous studies and notably
with the prediction by Baugh and Efstathiou (1993) from the deprojection
of the angular correlation function for the full APM galaxy survey. On
larger scales $\Xi(r)/r$ is as yet poorly determined due to the small
number of fields and their irregular geometry.

\begin{figure}
\centering 
\vspace*{5pt}
\parbox{\textwidth}{\epsfxsize=\textwidth \epsfbox{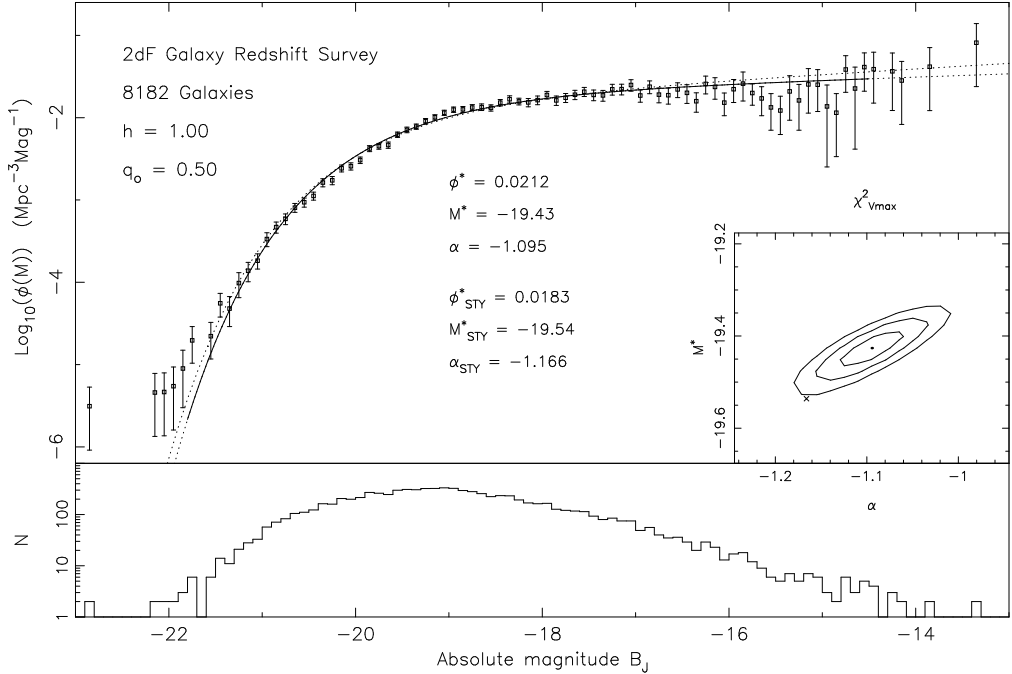}}
\vspace*{5pt}
\caption{A preliminary galaxy luminosity function from the 2dF survey
using mean K-corrections and normalised to the APM number counts. The
points are the $1/V_{max}$ LF, the solid curve is the $\chi^2$ fit of a
Schechter function to the points and the dotted curve is the STY fit of
a Schechter function. The parameters of these fits are given on the
figure. The inset shows the 1, 2 and 3-$\sigma$ contours of the $\chi^2$
fit in $M^*$ and $\alpha$; the cross marks the $M^*$ and $\alpha$
obtained from the STY fit. 
\label{fig:phi}}
\end{figure}

Although so far we have observed too few fields to effectively address
questions of large-scale structure, we do have a sufficiently large
sample of redshifts to begin to look at the properties of local
galaxies. Figure~\ref{fig:phi} shows the galaxy luminosity function for
the entire sample. This preliminary determination uses a single global
K-correction and is normalized by the number counts from the APM input
catalogue, so that the both the shape and normalization may change in
the final analysis. Note however that we are achieving a good
determination of the luminosity function 5 magnitudes below $L^*$ even
with this small subset of the full survey. The LF is generally well
represented by a Schechter function with $M^* \approx -19.5$, $\alpha
\approx -1.1$ and $\phi^* \approx 0.02$. 

\begin{figure}
\centering 
\vspace*{5pt}
\parbox{\textwidth}{\epsfxsize=\textwidth \epsfbox{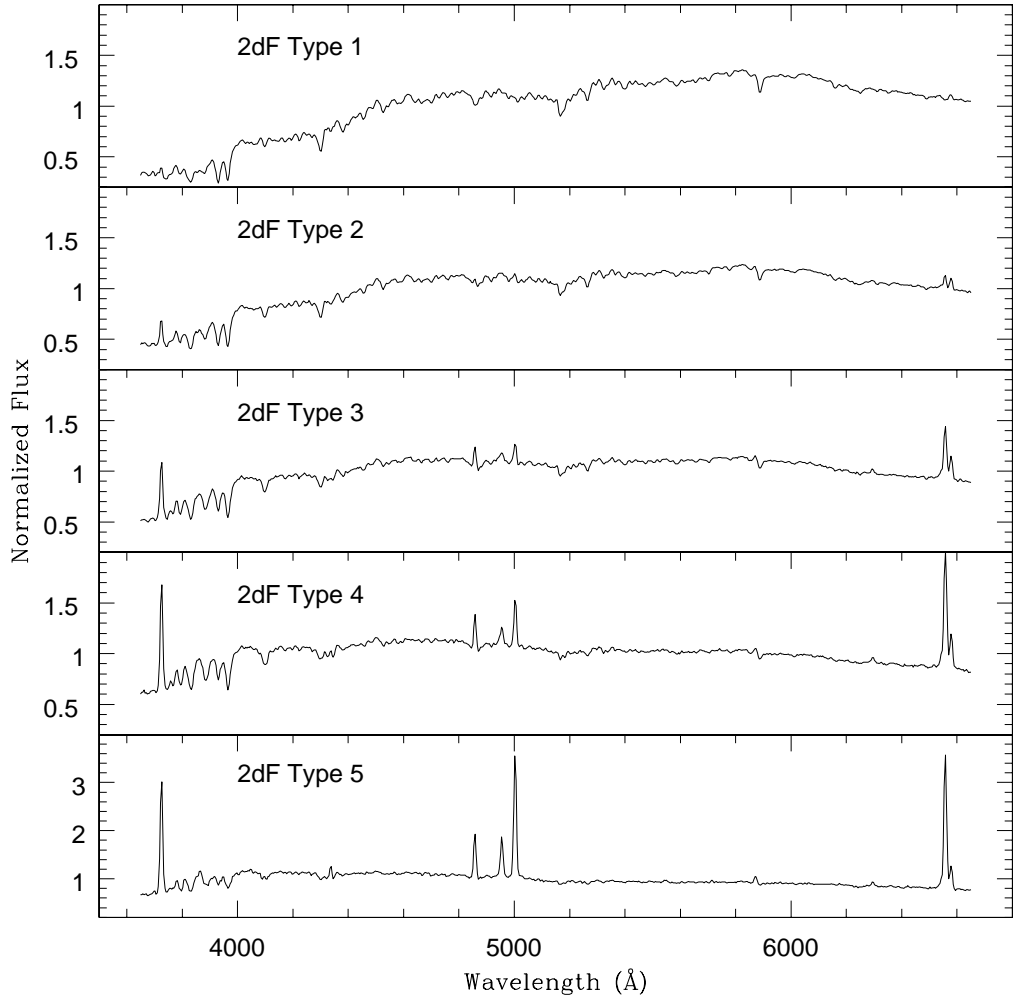}}
\vspace*{5pt}
\caption{The mean spectra corresponding to the five spectral types
defined from the principal component analysis of 3000 spectra.
\label{fig:means}}
\end{figure}

The crucial missing element in this analysis is a spectral
classification scheme, needed both in order to determine K-corrections
for individual galaxies and also in order to investigate the variation
in the luminosity function for different galaxy populations. As a first
step towards this goal, we have determined spectral types from a
principal component analysis for a subset of 3000 galaxies (Folkes
et~al.\ 1996). The galaxies were split into five spectral types based on
their projection in the plane of the first two principal components
(Folkes 1998). The mean spectra corresponding to each of these five
spectral types are shown in Figure~\ref{fig:means}. Type~1 corresponds
to a purely absorption-line galaxy spectrum and type~5 corresponds to a
strongly emission-line dominated spectrum, with the other types
intermediate.

\begin{figure}
\centering 
\vspace*{5pt}
\parbox{\textwidth}{\epsfxsize=\textwidth \epsfbox{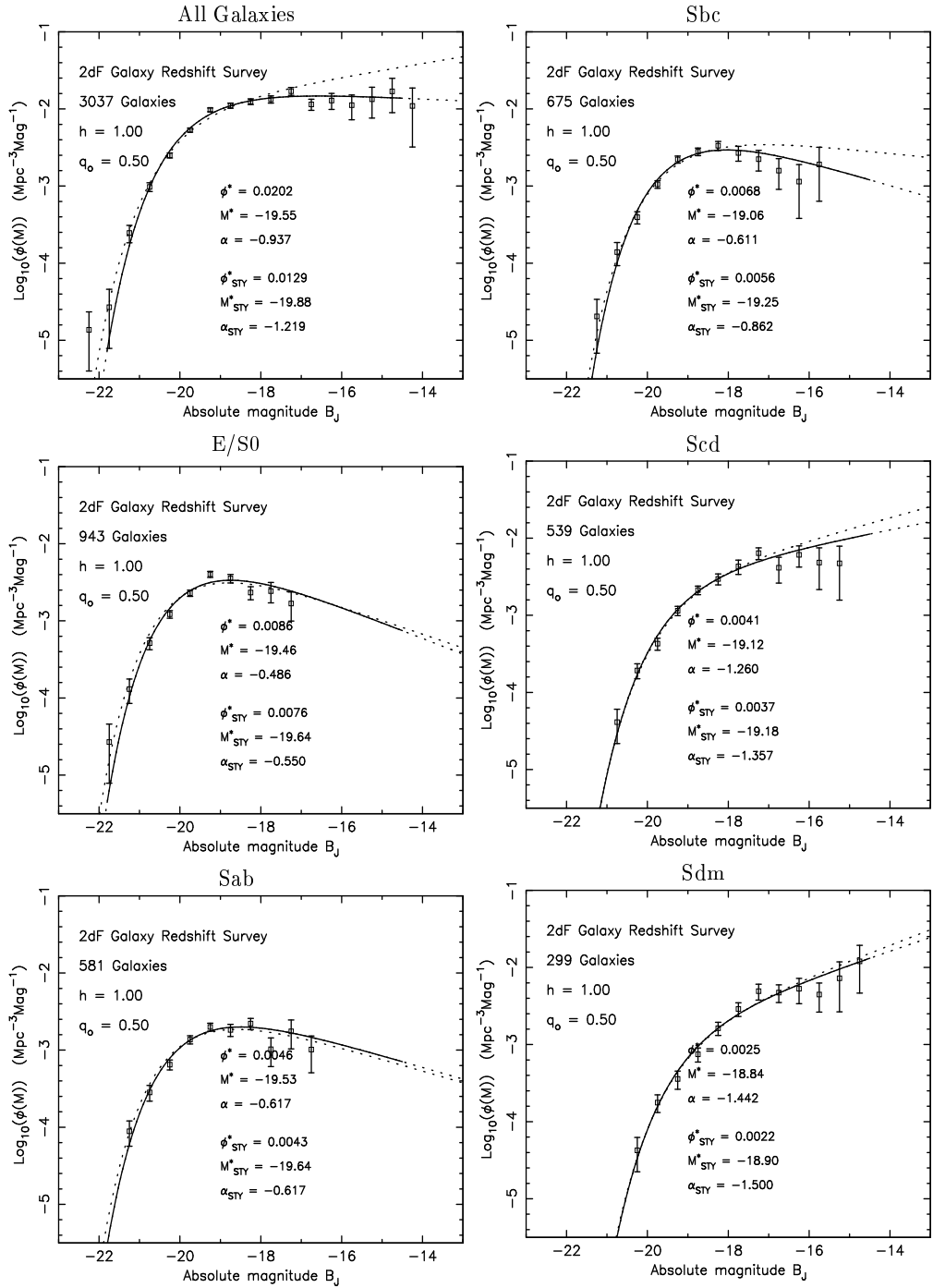}}
\vspace*{5pt}
\caption{The luminosity functions for the subsample of 3000 galaxies
with spectral types (and hence K-corrections) and for each of the
spectral classes individually.
\label{fig:lumfn}}
\end{figure}

The luminosity function for these 3000 galaxies with spectral types, now
with appropriate K-corrections, is shown in the top left panel of
Figure~\ref{fig:lumfn}. The remaining panels of the figure show the
luminosity functions for each type separately, and reveal a trend
towards fainter characteristic magnitudes $M^\ast$ and steeper faint end
slopes $\alpha$ as we move from `early' types (with absorption-line
spectra) to `late' types (with emission-line spectra). With the full
survey sample we will be able to refine this analysis in exquisite
detail, determining the variations in the luminosity function with both
spectral type and environment (i.e.\ local density) simultaneously.

\vspace*{10pt} 

Previous descriptions of the 2dF galaxy redshift survey and the 2dF
facility can be found in Colless (1997), Maddox (1998) and Colless and
Boyle (1998). Updates are posted on the WWW at {\tt
http://msowww.anu.edu.au/$\sim$colless/2dF/} and {\tt
http://www.ast.cam.ac.uk/$\sim$2dFgg/}.

\thebibliography{}
\item
Baugh C.M., Efstathiou G., 1993, MNRAS, 265, 145
\item 
Cole S., Hatton S., Weinberg D.H., Frenk C.S., 1998, MNRAS, in press
\item
Colless M.M., 1997, Wide Field Spectroscopy and the Universe, {\it Wide
Field Spectroscopy}, eds Kontizas M., Kontizas E., Kluwer, pp.227--240
\item
Colless M.M., Boyle B.J., 1998, in `Highlights of Astronomy', Vol.11, in
press
\item
Folkes S.R., 1998, PhD thesis, University of Cambridge
\item
Folkes S.R., Lahav O., Maddox S.J., 1996, MNRAS, 283, 651
\item
Glazebrook K., Offer A.R., Deeley K., 1998, ApJ, 492, 98
\item
Maddox S.J., 1998, in `Large Scale Structure: Tracks and Traces', World
Scientific, in press
\item
Maddox S.J., Efstathiou G., Sutherland W.J., Loveday J., 1990a, MNRAS,
242, 43{\sc p}
\item
Maddox S.J., Sutherland W.J., Efstathiou G., Loveday J., 1990b, MNRAS,
243, 692
\item
Maddox S.J., Efstathiou G., Sutherland W.J., 1990c, MNRAS, 246, 433
\item
Schlegel D.J., Finkbeiner D.P., Davis M., 1998, ApJ, 499, in press
\endthebibliography

\end{document}